\documentclass[prl,twocolumn,english,showpacs]{revtex4-1}
\usepackage{amssymb}
\usepackage{color}

\usepackage{graphicx}

\usepackage{bbold}
\usepackage{bm}
\usepackage{amsmath}
\usepackage{amsfonts}
\usepackage{amssymb}
\usepackage{braket}
\usepackage{units}

\usepackage{csquotes}
\usepackage{fancyhdr}

\begin{document}

\title{Simulating a Mott insulator using attractive interaction}

\author{M. Gall, C. F. Chan,  N. Wurz, and M. K\"ohl}
\affiliation{Physikalisches Institut, University of Bonn, Wegelerstra{\ss}e 8, 53115 Bonn, Germany}

\begin{abstract}
{We study the particle-hole symmetry in the Hubbard model using ultracold fermionic atoms in an optical lattice. We demonstrate the mapping between charge and spin degrees of freedom and, in particular, show the occurrence of a state with \enquote{incompressible} magnetisation for attractive interactions. Our results present a novel approach to quantum simulation by giving access to strongly-correlated phases of matter through an experimental mapping to easier detectable observables.} 
\end{abstract}

\maketitle Particles can be transformed into their anti-particles by a charge conjugation, and a symmetry upon such conjugation plays a crucial role in physics. For example, all fundamental forces, except for the weak interaction, obey such a symmetry \cite{Sozzi2008Book}. Although, historically, the particle-antiparticle symmetry was mainly studied in the context of elementary particles \cite{Wu1957,Christenson1964}, the concept can be extended to a wide range of problems. For example, in the low-energy domain of condensed matter physics, the natural anti-particle for an electron is a positively charged hole (absence of electron). Arguably, the most direct realization of particle-hole symmetry  occurs in  graphene, where the energy dispersion is a Dirac cone and hence particle-antiparticle symmetry becomes implicit \cite{DiVincenzo1984,GeimRMPGraphene}. More generally,  particle-hole symmetry leads to electrons and holes exhibiting the same physical properties such as effective mass, interaction and transport coefficients \cite{Kittel1987}. This impacts even  collective effects, such as the stability of the Higgs mode in BCS (Bardeen-Cooper-Schrieffer) superconductors \cite{Littlewood1981, Pekker2015, Behrle2018}.  Finally, it has been pointed out  that particle-hole symmetry can present a new pathway for the quantum simulation of strongly correlated materials \cite{Ho2009AttractiveHubbard}, an effect we are  exploring in this manuscript.

\begin{figure}[t]
	\includegraphics[width=0.48\textwidth]{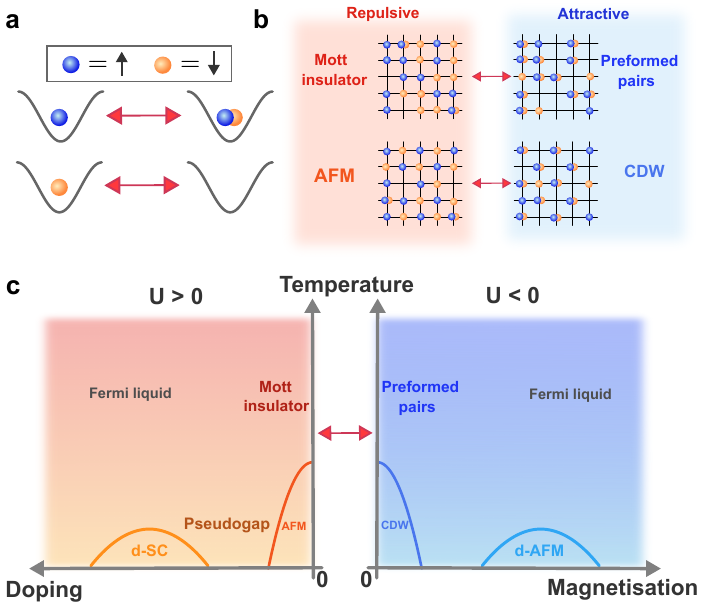}
\caption{Particle-hole transformation. \textbf{a}\, Mapping of site occupations. Under the transformation in Eq.~(\ref{PHtransformation}), a singly-occupied site with spin up (down) maps to a doubly-occupied (empty) site and vice versa. \textbf{b}\, Mapping of many-body states. The paramagnetic Mott insulator transforms into the preformed pairs phase, where attractive interactions favour the formation of pairs (doubles). Similarly, the spin ordering in the anti-ferromagnet (AFM) transforms into a charge ordering of the charge density wave (CDW). \textbf{c}\, Mapping of schematic phase diagrams. (left) The repulsive Hubbard model supports a wide range of strongly-correlated phases, which can be particle-hole transformed to phases in the attractive Hubbard model (right). We note that the role of doping $\tilde{n}$ interchanges with the role of spin-imbalance, i.e. magnetisation $m$.}
	\label{fig1}
\end{figure} 

For an ideal Fermi gas in a lattice, an exchange of particles to holes essentially maps occupied sites into empty sites, and vice versa. A particle-hole transformation in a spinful interacting system conjugates one of the spin components and is given by
\begin{equation}
	\begin{split}
		c_{i,\uparrow} & \rightarrow c_{i,\uparrow} \\
		c_{i,\downarrow} & \rightarrow (-1)^{i_x+i_y} c^\dagger_{i,\downarrow}, \\
	\end{split}
	\label{PHtransformation}
\end{equation}
where $i = (i_x,i_y)$ denotes the two-dimensional lattice site index and $c^{\dagger}_{i,\sigma} (c_{i,\sigma} )$  the fermionic creation (annihilation) operator at site $i$ with spin $\sigma = \uparrow, \downarrow $.  As shown in Fig.~1a, this transformation maps the singly-occupied spin-up and spin-down states (singles) to doubly-occupied (doubles) and empty sites (holes), respectively. Hence, the transformation interchanges the spin with the density sector, e.g. a spin-imbalance transforms into an imbalance of doubles and holes, equivalent to a change in density. 
 

Applying the mapping of local occupations, a paramagnetic Mott insulator with predominant occupation of one particle per site transforms into a phase with local preformed pairs. Moreover, an anti-ferromagnetically ordered state (AFM) maps into a charge density wave (CDW) \cite{Ho2009AttractiveHubbard,EsslingerReview2010}, as depicted in Fig.~\ref{fig1}b. However, the consequences of this transformation are more profound than the interchange of the local site occupations, because even the equilibrium many-body states of the particle-hole symmetric Hamiltonian are mapped. Therefore, correlation functions and thermodynamic observables also have a corresponding counterpart. For example, the direct phase mapping in the Hubbard model is shown in Fig.~\ref{fig1}c. This connection gives access to phases which may be difficult to detect by direct means.  For instance, well-developed experimental techniques for the spin sector, such as measuring momentum-resolved spin correlations \cite{Wurz2018}, can be used to reveal the physics of density-ordered states, and vice versa.

In this work, we realise a single-band Hubbard model, the Hamiltonian of which is given by
\begin{equation}
H = -t \sum_{\langle i,j \rangle, \sigma}  c^\dagger_{i,\sigma} c_{j,\sigma} + U \sum_{i} \tilde{n}_{i,\uparrow}  \tilde{n}_{i,\downarrow}  - \sum_{i,\sigma} \mu_{\sigma} \tilde{n}_{i,\sigma}.
\label{HubbardHamiltonian}
\end{equation}
Here, $t$ is the nearest-neighbour tunnelling, $U$ is the interaction parameter, $c^\dagger_{i,\sigma} c_{i,\sigma} = \tilde{n}_{i,\sigma}  + 1/2$ is the fermion number operator and $\mu_{\sigma}$ is the chemical potential for spin $\sigma$. To separate the effects of doping and spin-imbalance, the third term in Eq.~(\ref{HubbardHamiltonian}) can be rewritten as $\sum_{i,\sigma} \mu_{\sigma} \tilde{n}_{i,\sigma} = \mu \sum_{i,\sigma} \tilde{n}_{i,\sigma} + h \sum_{i} \left( \tilde{n}_{i,\uparrow} - \tilde{n}_{i,\downarrow} \right)$, where $\mu = (\mu_{\uparrow} + \mu_{\downarrow})/2$ is a spin-independent chemical potential and $h = (\mu_{\uparrow} - \mu_{\downarrow})/2$ is an effective Zeeman (magnetic) field. At chemical potential $\mu = 0$, there is on average one atom per lattice site (half-filling) and the density doping $\tilde{n} = n_{\uparrow} + n_{\downarrow} - 1$ is zero. Changing the chemical potential to $\mu \neq 0$ leads to finite doping, which is equivalent to the difference of doubles and holes. 
In a similar fashion, the Zeeman field $h$ determines the difference of the spin densities, resulting in a finite magnetisation $m = n_{\uparrow}- n_{\downarrow}$ away from the spin-balanced case at $h = 0$. Under the particle-hole transformation of Eq.~(\ref{PHtransformation}), the interaction term is inverted $U \rightarrow -U$ and the generalised forces $\mu$ and $h$ exchange their roles. Hence, the particle-hole symmetry provides a tight connection between the phases with repulsive and attractive interactions. In Fig.~\ref{fig1}c, we display the phase diagram of the Hubbard model and its corresponding phases under the particle-hole transformation. Here, the mapping of site occupations results in the transformation of the magnetisation $m$ to the density doping $\tilde{n}$
.  Even though this link has long been known \cite{Giamarchi2004,Ho2009AttractiveHubbard}, it has mostly been discussed in a pedagogical fashion.

\begin{figure}[t]
\includegraphics[width=0.48\textwidth]{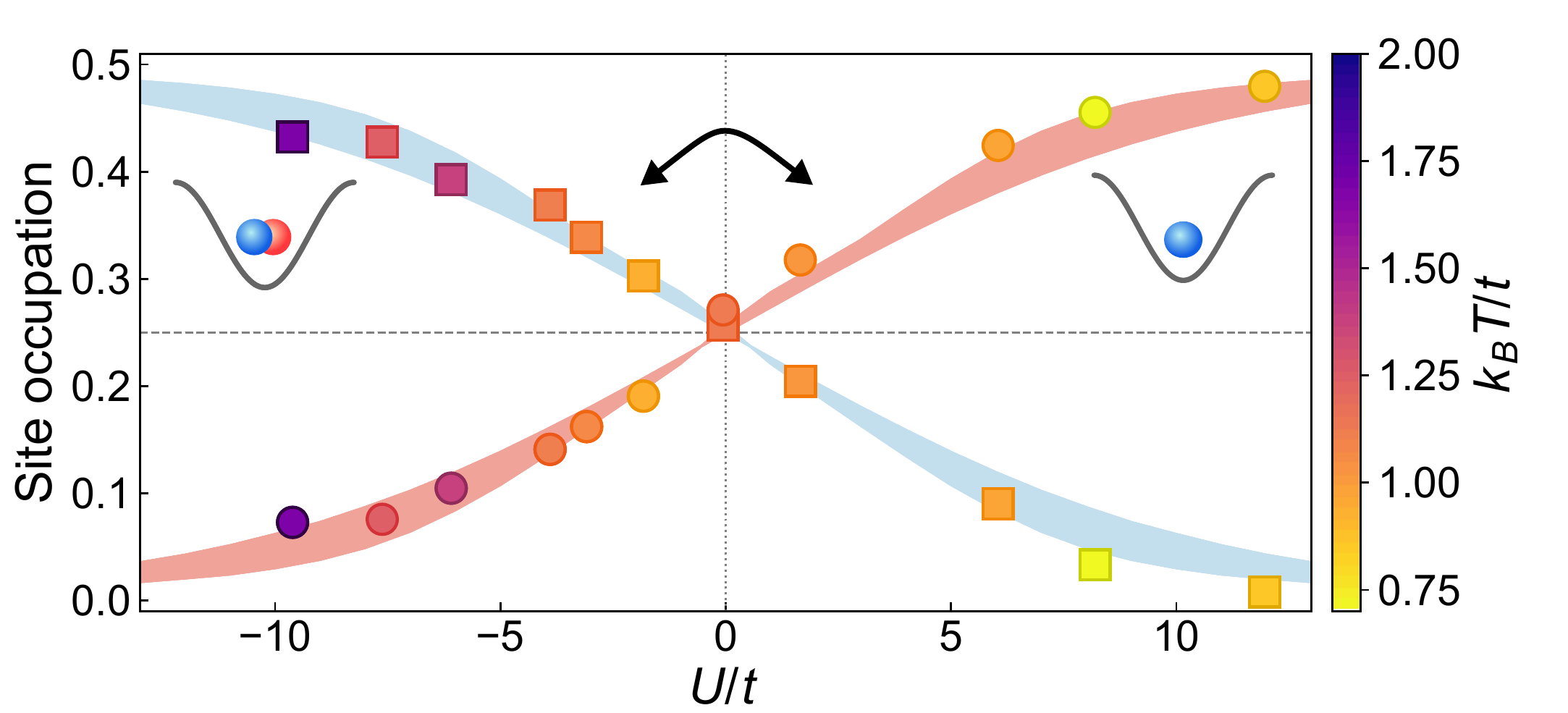}
\caption{Particle-hole symmetry for various interaction strengths. Doubles (squares) and singles (circles) site occupation at $\mu=h=0$ versus $U/t$. For each interaction, we show the lowest temperature reached in our data sets. The red (blue) shaded regions show the results of the DQMC numerical calculations over \mbox{$0.8 \leq k_B T/t \leq 2$.}}
\label{fig2}
\end{figure}

We investigate both the repulsive ($U>0$) and attractive ($U<0$) two-dimensional Hubbard model using a two-component ultracold Fermi gas of potassium atoms loaded into an optical lattice. To realise the two-dimensional Hubbard model, we start from a degenerate two-component Fermi gas of $^{40}K$ atoms in the two lowest hyperfine ground states, $\ket{\downarrow}= \ket{F = 9/2, m_F = -9/2}$ and $\ket{\uparrow} = \ket{F = 9/2, m_F = -7/2}$.  The atoms are loaded into the lowest band of an anisotropic three-dimensional optical lattice potential with strongly suppressed tunnelling along the vertical direction. Within the horizontal planes an almost square lattice potential is applied with a lattice constant of $a = 532\,\mathrm{nm}$ and a depth of $6.0(1) E_{rec}$ or $5.8(1) E_{rec}$ resulting in a tunnelling amplitude of $t/h = 224(6)\,\mathrm{Hz}$ or $235(7)\,\mathrm{Hz}$, respectively. Here, $E_{rec} = \frac{h^2}{8ma^2}$ is the recoil energy, $m$ is the atomic mass and $h$ is the Planck constant. The interaction strength $U$ is varied by tuning a vertically aligned magnetic bias field in the vicinity of the Feshbach resonance located at $202\,$G. After preparing the atoms in the lattice, we freeze their positions by quickly (within $1\,$ms) ramping the horizontal lattice depth to a value of $60\,E_{rec}$, thereby suppressing dynamics in all directions. This is followed by radio frequency (RF) tomography in the presence of a vertical magnetic field gradient in order to address a single two-dimensional layer.

Subsequently, we implement two different detection protocols, corresponding to the measurements in the charge and spin sectors. For the detection of singles and doubles in a single repulsive (attractive) Hubbard model, we perform RF spectroscopy at a homogeneous magnetic field of $180\,$G ($213\,$G). Due to the interaction shift between the states, we spectroscopically separate the singles and doubles which allows subsequent detection. For the spin-resolved measurement, we make use of the spin-changing collision between $\ket{F = 9/2, m_F = -9/2}$ and $\ket{F = 9/2, m_F = -3/2}$ to remove doubly-occupied sites. Finally, absorption images of singles/doubles or spin-up/spin-down singles are taken \cite{Cocchi2016,Drewes2017}. We exploit the knowledge of the harmonic confinement to extract the chemical potential landscape $\mu(x,y)$ under the assumption of the local density approximation. In the preparation of our system, we introduce a variable imbalance of the two spin components, which leads to a global, finite $h$. By tuning the magnetic field close to an s-wave Feshbach resonance, we realise a wide range of interactions $-10 < U/t < 12$. We compare our experimental data to determinant quantum Monte-Carlo (DQMC) simulations \cite{Varney2009}, in order to extract the global temperature $T$, the effective Zeeman field $h$ and the chemical potential $\mu$ in the centre of the trap.

A straightforward starting point to study the consequences of the particle-hole symmetry is at half-filling ($\mu = 0$) and for a spin-balanced cloud ($h=0$), since the mapping $\mu \leftrightarrow h$ is automatically fulfilled.  
In Fig.~2, we show the site occupation of singles (spin-up) and doubles over a wide range of interactions. 
In the non-interacting case at $U = 0$, doubles are energetically neither favoured nor suppressed compared to singles or empty sites. Therefore, at half-filling, all possible states are equally likely at $1/4$ each. By gradually increasing the repulsive interaction strength ($U>0$), doubles become energetically less favourable. Therefore, we expect a decrease of doubles accompanied by an increase of singles. The opposite takes place for attractive interactions ($U<0$), where doubles are favoured and singles are suppressed.
In Fig.~2, the particle-hole symmetry manifests itself as a precise mirror symmetry of the site occupations of spin-up singles and doubles about $U = 0$.

\begin{figure}[h]
	\includegraphics[width=0.47\textwidth]{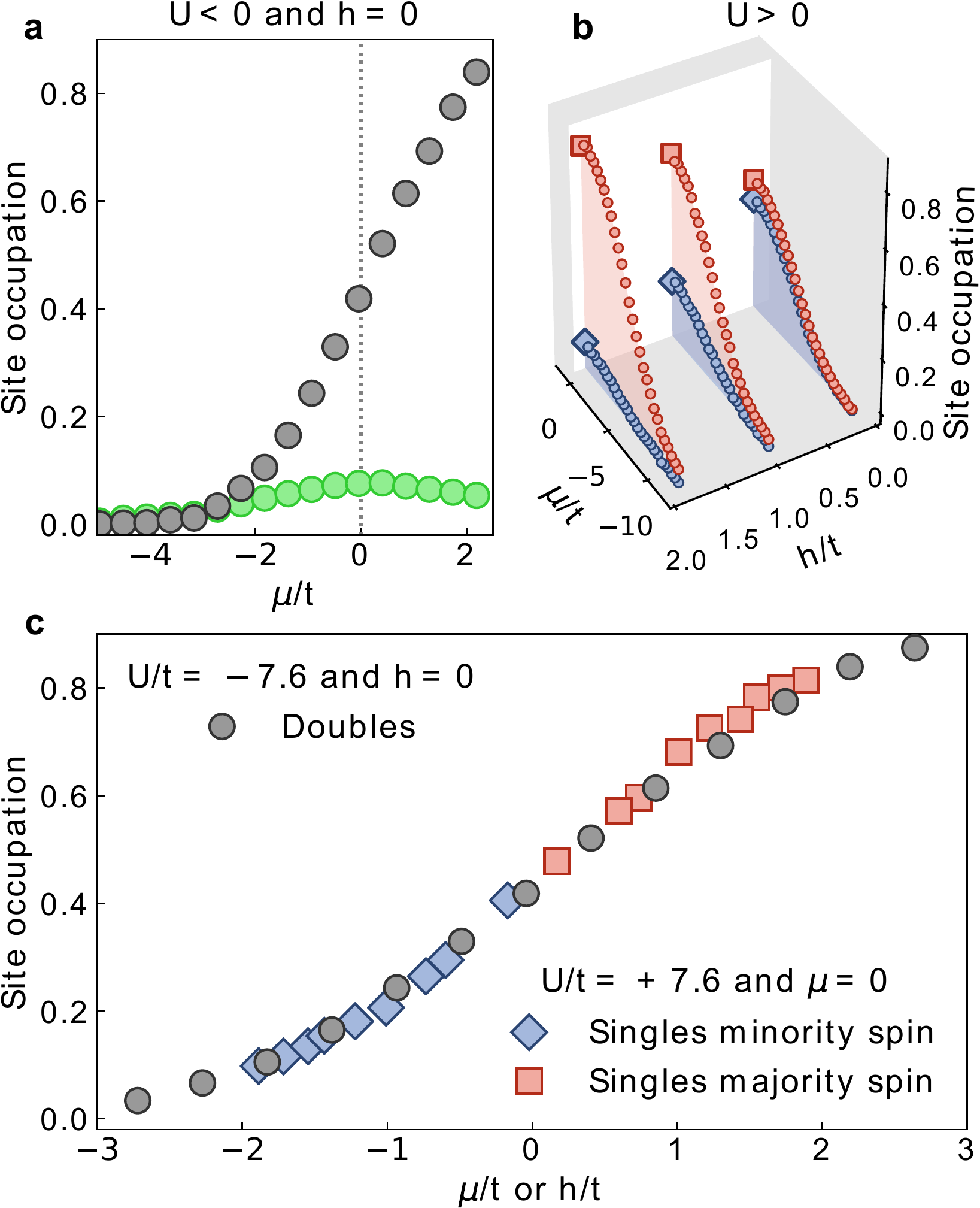}
	\caption{Interchanging the role of effective Zeeman field with the chemical potential. 
		\textbf{a}\, For the attractive, spin-balanced ($h=0$) Hubbard model ($U/t = -7.6$) we show site occupations of singles (green circles) and doubles (black circles) as a function of $\mu$.
		\textbf{b}\, For the repulsive Hubbard model ($U/t = 7.6$) with a finite spin-imbalance ($h \neq 0$) we show site occupation of the majority/minority spin component (red/blue markers) as a function of both $h$ and $\mu$. The white plane indicates half-filling $\mu = 0$. The data points intersecting this plane are highlighted with squares and diamonds.
		\textbf{c}\, Combining data sets from a and b  within a similar temperature range ($T/t \approx 1.5$) leads to a  collapse of the data onto the same curve.  	}
	\label{fig3}
\end{figure}

 
Next, we generalise to arbitrary values of $\mu$ and $h$, the role of which should interchange when transforming $U \rightarrow -U$. In Fig.~3a, we show the equation of state of singly- and doubly-occupied sites for an attractive Hubbard model with $U/t= -7.6$. The attractive interaction leads to an excess of doubly-occupied sites near the center of the trap (large chemical potential). Calibration of the half-filling point $\mu =0$ is provided by the peak in the singles density. While the chemical potential $\mu$ imbalances the populations of doubles and holes when moving away from half-filling, the effective Zeeman field $h$ takes on a similar role if we consider the spin-up and spin-down populations instead. In order to investigate this, we switch to repulsive interactions ($U/t = 7.6$), introduce a global imbalance between spin-up and spin-down states, and change our detection routine to simultaneously probing singles of both spin components \cite{Drewes2017}. In Fig.~3b, the measured site occupations of spin-up and spin-down singles are plotted as a function of $\mu$ for different global Zeeman fields $h$. We extract the site occupations at $\mu = 0$, in order to compare to the spin-balanced data ($h=0$) with attractive interactions.
In Fig.~3c, we combine the previously discussed measurements with attractive and repulsive interactions. We plot the site occupation of doubles ($U<0$) and spin-up singles ($U>0$) together using the same axis scale for the effective Zeeman field $h$ and the chemical potential $\mu$. Since flipping the sign of $h$ interchanges the populations
of spin-up and spin-down, we can use the site occupations of the spin-down singles when inverting the Zeeman field $h$. 
The excellent agreement of our data, when interchanging the role of $\mu$ and $h$, proves the validity of the particle-hole symmetry in our quantum simulator.

\begin{figure}[t]
\includegraphics[width=0.47\textwidth]{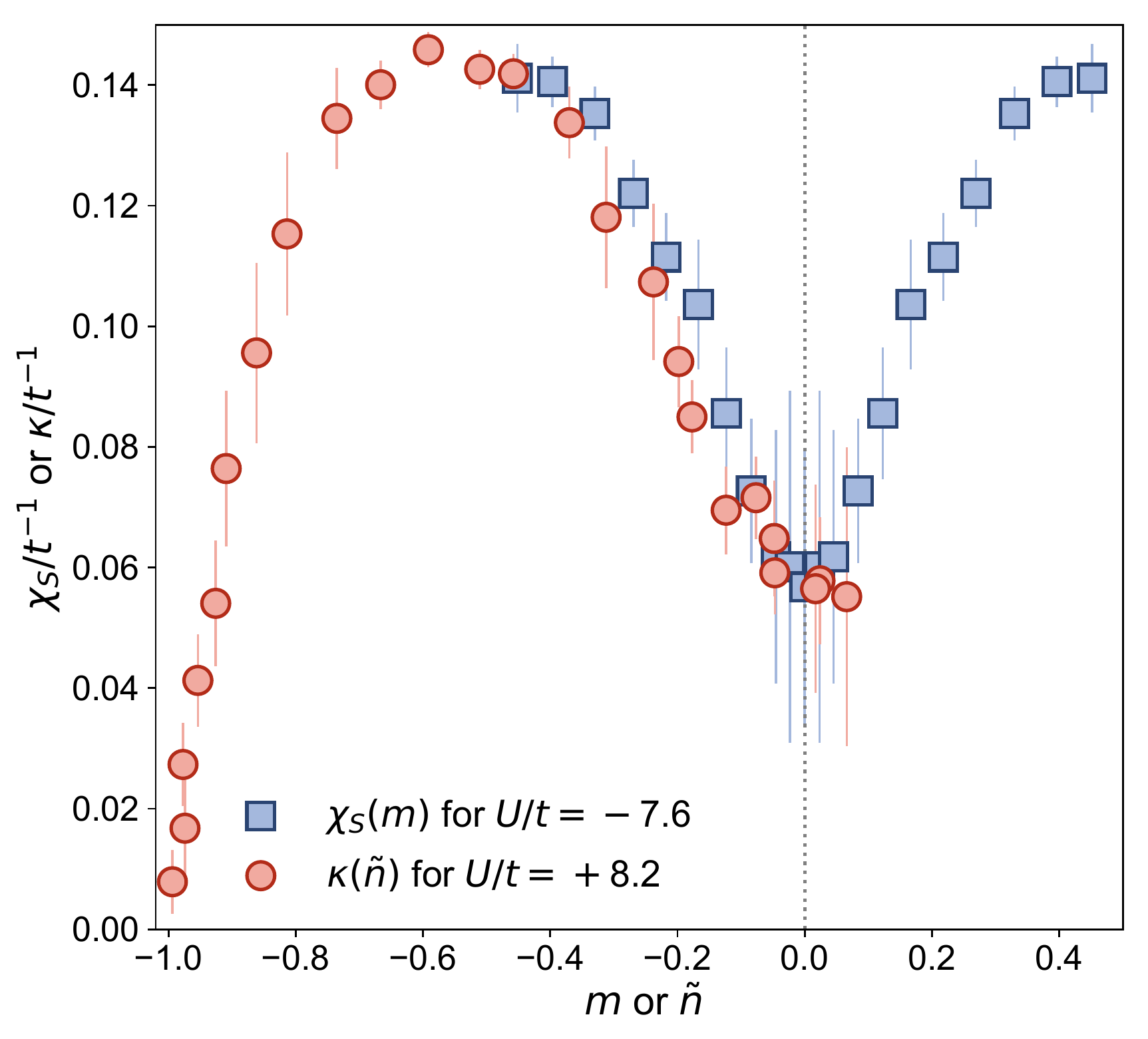}
\caption{Observing the Mott-like incompressibility with attractive interactions. Spin susceptibility $\chi_s$ (blue squares) at half-filling and attractive interactions compared to the compressibility $\kappa$ (red circles) in a spin-balanced system with repulsive interactions \cite{Cocchi2016}. We measure the magnetisation $m$ at half-filling for a large range of the effective Zeeman field $h$ up to $4.7\, t$. The spin susceptibility $\chi_s$ is obtained by performing the numerical derivative $\partial{m}/\partial{h}$. The error bars represent the standard errors.}
\label{fig4}
\end{figure}

Finally, we extend the investigation of the particle-hole symmetry to the phase diagram of the Hubbard model. The Mott insulator has been studied in detail over the past years  \cite{Joerdens2008,Schneider2008,Greif2015,Cheuk2016,Cocchi2016}. At low temperatures, the system favours singly-occupied sites over doubles and holes due to the repulsive on-site interaction. Therefore, around half-filling, the isothermal compressibility $\kappa = \partial \tilde{n} / \partial \mu$ is low, signaling that it is energetically unfavourable to accommodate more particles when raising the chemical potential. This is the hallmark of the insulating nature of this state \cite{Imada1998,Joerdens2008,Schneider2008, Cocchi2016}.

The corresponding phase for attractive interactions is dominated by preformed pairs, which suppress the local magnetisation $m$. In contrast to the repulsive case, where the system opposes the formation of doubles, it is now unfavourable to break them, even when the spin populations get slightly imbalanced. This is observed as a minimum of the static spin susceptibility $\chi_s = \partial m / \partial h$ at $m=0$. However, even beyond this qualitative argument, the particle-hole transformation provides the exact mapping between the two correlation functions $\kappa$ and $\chi$, which should precisely equal each other. We display our measurement of $\kappa(\tilde{n})$ and $\chi_s(m)$ in Fig.~\ref{fig4} and despite comparing systems with opposite interaction strengths and different observables, we achieve strikingly good agreement. Therefore, we demonstrate that we can use the particle-hole symmetry to simulate the Mott phase with a spin-imbalanced system and attractive interactions.

Having proven the direct mapping of the spin to the density sector, our work opens a promising road to investigate yet unexplored phases using quantum simulators. The principle demonstrated here allows for the simulation of phases in an experimentally better accessible parameter regime. 
For example, whether or not a d-wave superconducting phase exists in the doped repulsive Hubbard model is still debated due to the lack of suitable experimental and numerical techniques \cite{EsslingerReview2010}. 
The particle-hole symmetry maps the superconducting d-wave pair correlations  in momentum space
 to a spin-imbalanced attractive system with d-wave AFM correlations in real space 
\cite{Ho2009AttractiveHubbard} which could be observed with excellent spatial detection capabilities \cite{Mazurenko2016,Drewes2017,Hart2015,Guardado-Sanchez2018,Cheuk2016b,Boll2016}.

This work has been supported by BCGS, the Alexander-von-Humboldt Stiftung, ERC (grant 616082), DFG (SFB/TR 185 project B4) and Stiftung der deutschen Wirtschaft.

\section{Supplementary material}

\subsection{DQMC simulation}

The DQMC simulations are performed using the Quantum Electron Simulation Toolbox (QUEST) Fortran package \cite{Varney2009}. Simulations are performed for a homogeneous $8 \times 8$ square lattice with a wide range of interactions $-12 \leq U/t \leq 12$, with $1000$ warm-up sweeps and $10000$ measurement sweeps, and the number of imaginary time slices is set to $24$. For the spin-balanced case, the chemical potential $\mu/t$ is varied through a range of -20 to 20, and temperature $T/t$ is scanned over a range of 0.4 to 4. Additionally, for the spin-imbalanced case, simulations are performed using spin-dependent chemical potentials, whose difference $2h = (\mu_\uparrow - \mu_\downarrow)$ is scanned from 0 to 30. However, for attractive interactions and finite $h$ the simulation suffers from a sign problem. Thus, in order to quantitatively compare our measurement in the  attractive spin-imbalanced case, repulsive data is transformed under the particle-hole symmetry as stated in Eq.~(1).

\end{document}